# On the Off-chip Memory Latency of Real-Time Systems: Is DDR DRAM Really the Best Option?

Mohamed Hassan
*University of Guelph*, Canada, mohamed.hassan@uoguelph.ca
*Intel Corporation*, Canada, mohamed1.hassan@intel.com

*Abstract*—**Predictable execution time upon accessing shared memories in multi-core real-time systems is a stringent requirement. A plethora of existing works focus on the analysis of Double Data Rate Dynamic Random Access Memories (DDR DRAMs), or redesigning its memory to provide predictable memory behavior. In this paper, we show that DDR DRAMs by construction suffer inherent limitations associated with achieving such predictability. These limitations lead to 1) highly variable access latencies that fluctuate based on various factors such as access patterns and memory state from previous accesses, and 2) overly pessimistic latency bounds. As a result, DDR DRAMs can be ill-suited for some real-time systems that mandate a strict predictable performance with tight timing constraints. Targeting these systems, we promote an alternative off-chip memory solution that is based on the emerging Reduced Latency DRAM (RLDRAM) protocol, and propose a predictable memory controller (RLDC) managing accesses to this memory. Comparing with the state-of-the-art predictable DDR controllers, the proposed solution provides up to $11\times$ less timing variability and $6.4\times$ reduction in the worst case memory latency.**

## I. Introduction

With the Internet-of-Things (IoT) revolution, real-time systems are unprecedentedly becoming ubiquitous in our daily life. Examples include healthcare devices, automotive, and smart power grids. In these systems, a failure can result in severe consequences such as loss of lives. This failure is possible not only by incorrect functionality, but also by violating temporal requirements. Accordingly, a detailed worst-case execution time (WCET) analysis (statically or experimentally) of a real-time task's execution is necessary to ensure satisfying the task's temporal requirements. Hardware components have to follow a predictable behavior to allow for this analysis. Unfortunately, conventional computing systems are not designed to be predictable. Numerous architectural optimizations such as deep pipelines, branch prediction, and aggressive reordering aim to provide high performance at the expense of immense timing variability. Since failures in real-time systems have to be avoided at all costs, hardware architecture has to be reconsidered to account for predictability in the first place.

Considering off-chip main memory, which is a critical component of most computing systems [1], we find that Double Data Rate Dynamic Random Access Memories (DDR DRAMs) or simply (DDRx[1]) are the most used nowadays. This is because they provide a low-cost, a high-capacity, and a high-bandwidth solution for performance oriented systems. Despite the name, DDRx DRAMs in reality do not provide random access. Their access latency varies notably based on many factors such as access patterns, transaction type (read or write), and the DRAM state from previous accesses. Moreover, to further boost DDRx performance and reduce access latency, memory controllers usually employ complex optimizations such as multiple reordering levels, prioritizations, and adaptive policies [2]. These memory controller optimizations along with the variability of DDRx's access latency result in highly pessimistic worst case latency (WCL) [3], [4], which encumbers the deployment of DDRx in real-time systems. To address this challenge, researchers proposed redesigning the memory controller to provide predictable access to DDRx memories [5]–[19]. This approach helps in reducing the interference latency amongst requests belonging to different tasks or processing elements (PEs). Nonetheless, it does not reduce the variability in the access latency of the DDRx itself. This access latency is inherently bounded by the physical characteristic of the DDRx chips and is enforced by the JEDEC standard constraints [20]. DDRx internal circuitry by construction targets average-case performance with complex interactions between DDRx commands and more than 20 timing constraints [17]. As we show in this paper, this inherent variability greatly affects the resulting WCL even when adopting one of the aforementioned predictable memory controllers in the system. As a consequence, we believe revolutionary solutions have to be devised to provide more predictable memory behavior with lower WCL for real-time systems.

In this paper, 1) we thoroughly study the variability of DDRx's access latency and expose its limitations with regard to real-time systems (Section III). 2) Motivated by these limitations, we explore alternatives to DDRx memories. Namely, we promote the adoption of Reduced Latency DRAM (RLDRAM) in real-time systems. RLDRAM is an emerging DRAM protocol that is currently led by Micron [21] and provides predictable behavior with lower access latency compared to DDRx protocols. We illustrate how RLDRAM can provide a considerable reduction in WCL as well as less access variability, which establishes the motivation towards adopting RLDRAM in real-time systems (Section IV). 3) To enable this adoption, we propose RLDC: a memory controller design that predictably manages accesses to the RLDRAM. We also conduct timing analysis that provides an upper bound on the latency suffered by any memory request upon accessing the RLDRAM using RLDC (Section V). 4) To show the

[1] we use the letter x since observations we make in this paper are generic for any DDR protocol (DDR2, DDR3, DDR4, etc.).

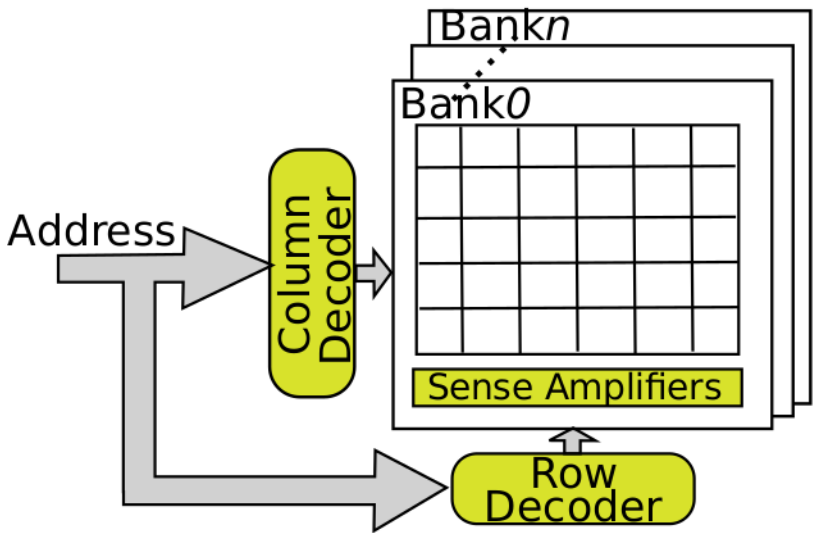

Fig. 1: DRAM architecture.

effectiveness of the proposed solution, we compare with eight of the state-of-the-art predictable DDRx controllers using representative benchmarks from the automotive domain. Results show that the proposed solution provides up to $6.4\times$ reduction in WCL and $11\times$ less latency variability (Section VI).

## II. Related Work

**Predictable DDRx solutions**. Existing works focus on providing predictability to real-time tasks upon accessing DDRx main memories (e.g. [3]–[19], [22], [23]). These efforts follow two major directions. The first direction is to analyze existing memory controllers used in conventional high-performance systems to upper bound the latency suffered by any request upon accessing DDRx main memory [3], [4], [22]. Following a similar direction, [23] targets to bound DRAM interference in conventional platforms by enforcing bank partitioning at the operating system level. As aforementioned, these commodity controllers target to increase average-case performance through complex optimizations at the expense of predictability. Consequently, the provided bounds by these approaches are pessimistic, which entails them ill-suited for real-time systems with tight timing requirements. The second direction is to entirely or partially redesign the memory controller to account for predictability (e.g. [5]–[19], [24], [25]). A comparative study that highlights strengths and limitations of some of these controllers is proposed in [26]. Although this approach reduces latency variabilities due to delay interferences among requests of different tasks, it suffers from the two main drawbacks. 1) It still suffers from high WCLs due to the complex interactions between DDRx commands. 2) It can not address the variability in the access latency of the DDRx chips. We show that access latency variability in DDRx memories severely affects the predictability of memory requests. We address this problem by promoting the deployment of RLDRAM emerging memories in real-time systems. We provide a predictable RLDRAM memory controller with tighter latency bounds and less variability compared to these DDRx controllers.

**RLDRAM**. RLDRAM memory is originally targeted at high-speed routers [27] and network processing [28]. Researchers also envisioned the usage of RLDRAM as a low-latency memory module in a heterogeneous memory system to increase overall memory performance [29], [30]. To our best knowledge, we are the first to investigate the usage of RLDRAM in real-time systems.

## III. DDRx Limitations

We first introduce the basics of the DDRx protocol. Then, we study its limitations with regard to providing predictability.

### A. DDRx DRAM Basics

Figure 1 illustrates the basic structure of DDRx DRAMs. They consist of an array of memory cells arranged as *banks*. Cells in each bank are organized in *rows* and *columns*. Multiple banks can construct a logical entity called a *rank*. Each bank has sense amplifiers, which also cache the most-recently accessed row in each bank (known as *row buffer*). As Figure 1 shows, the address bits are split into two segments. One segment is used by the row decoder to determine the requested row, and the other is used by the column decoder to determine which column to access within this row. DDRx memories use a multiplexed address mode such that these two segments are provided to the memory in two steps. First, the row address is provided to activate the requested row. Then, in a later cycle, the remaining address is provided to index specific column(s) in the activated row.

Although address multiplexing reduces the pin count of the DDRx chip (and hence reduces its cost), it increases the access latency as one memory access is now split into multiple stages. Namely, one access to a DDRx memory can comprise a maximum of three stages: 1) precharge, 2) activate, and 3) read or write. All these stages have to be orchestrated by an on-chip memory controller through issuing memory commands. A precharge command (P) writes back the data in the row buffer into DRAM cells. It is needed if the access is to a row different than the one in the row buffer. An activate command (A) fetches the requested row from the cells into the row buffer. Read/Write commands (R/W) conduct the requested memory operation. The controller also needs to periodically issue a REF command to negate the charge leakage from the capacitors that store the data. The effect of REF on predictability is deterministic and is limited to around 2% of task's memory latency [3]. In addition, the majority of the DDRx controllers do not incorporate this effect on their analysis as they conduct a per-request analysis, while the REF effect should be incorporated in WCET analysis at the task level to avoid pessimism [12], [26]. For these reasons, we do not consider the REF command in this paper.

All commands have strict timing constraints that are dictated by the JEDEC standard [20] and must be satisfied by all memory controller designs. Table I tabulates the most relevant timing constraints for DDR3-1600 DRAM. It is worth noting that, in addition to increasing the access latency, the aforementioned three stages (precharge, activate, and read/write) lead to high variability in access latency. This is because one request can consist of one, two, or three stages based on the memory state as follows. 1) If the request targets a row that already exists in the row buffer (denoted as *open row*), it only consists of a single stage and the controller issues only either a R or W command based on the request type. 2) If the request targets a bank that is already precharged (i.e. does not have an open row in the row buffer), it consists of two stages: the

TABLE I: JEDEC Timing Constraints [20].

| Parameter | Delay Description | Cycles |
|---|---|---|
| $tRCD$ | A to R/W | 10 |
| $tCCD$ | R to R or W to W (same rank) | 4 |
| $tRL$ | R to start of data transfer | 10 |
| $tRP$ | P to A | 10 |
| $tWL$ | W to start of data transfer | 9 |
| $tRTW$ | R to W | 6 |
| $tRTP$ | R to P | 5 |
| $tWTR$ | End of data transfer of W to R | 5 |
| $tWR$ | End of data transfer of W to P | 10 |
| $tRAS$ | A to P | 24 |
| $tRC$ | A to A (same bank) | 34 |
| $tRRD$ | A to A (diff bank in same rank) | 4 |
| $BL/2$ | Data bus transfer | 4 ($BL8$)* |
| $tRTRS$ | Rank to rank switch | 1 |

*BL is the burst length, which indicates the number of data beats to be transfered by one access.

activate stage to bring the row to the row buffer in addition to the read/write stage. We say in this case that the request targets a *closed row*. 3) Finally, if the request targets a bank that has an open row in the row buffer that is different than the targeted row by the request, it needs all the three stages and the controller in this case issues three commands on behalf of that request: P, A, and then R or W. We say in this case that the request is targeting a *conflict row*.

Finally, the data bus of the DDRx is bidirectional: same wires are used to both read from and write to the DRAM. This also increases access latency since the data bus needs to switch from read to write or vice versa. For instance, in DDR3-1600 devices, the R-to-W switching delay is 6 cycles, while the W-to-R delay is 18 cycles.

### B. Predictability Considerations

Since predictability is of utmost importance in real-time systems, we investigate in details the effect of DDRx's access latency variability on predictability. Predictability has different definitions in the real-time literature. One important measure of system predictability is the relative difference between best- and worst-case execution times (or latencies in case of memories) [31]. The latency of a memory request can be anywhere between the best-case latency (BCL) and WCL. To differentiate between the effects from the access protocol of the DRAM and the effects from the scheduling techniques of the controller, we define two latency components: memory access latency (Definition 1), and total memory latency (Definition 2).

*Definition 1:* **Memory Access Latency** of a request to the DRAM is measured from the time stamp when the request is elected by the scheduling mechanisms of the controller to be sent to the memory (i.e. it is at the head of the scheduling queue) until its data starts the transfer on the data bus.

*Definition 2:* **Total Memory Latency** of a request to the DRAM is measured from its arrival at the memory controller until the start of its data transfer.

Memory access latency accounts only for time consumed by the request itself to perform the access including delays suffered due to the current state of the memory (i.e. remaining timing constraints from previously issued requests). In contrast, total memory latency accounts for delays due to both arbitration decisions at the memory controller as well as the access latency to the DRAM. Since the target of this subsection is to study the predictability of the DRAM device itself and not the controller, we eliminate any scheduling-specific effects induced by the controller. Therefore, we focus only on the access latency as per Definition 1. The controller's scheduling effects on predictability is discussed in Section III-C. To obtain best- and worst-case DRAM access latencies, we study all possible access scenarios and their corresponding access latency. Then, to quantitatively measure the DRAM predictability, we define the term variability window in Definition 3.

*Definition 3:* **Variability Window** of a latency component, $VW$, is a measure of the possible variations in the value of this component and is computed as the percentage increase from the best ($BCL$) to the worst case latency ($WCL$) values of this component.

$$\text{VW} = \frac{WCL - BCL}{BCL} \times 100 \quad (1)$$

In DDRx memories, many factors determine the access latency of the request such as 1) the DRAM state from previous accesses, 2) the data bus direction based on the type of the current and previous transactions (read or write), and 3) the request address (i.e. targeted rank, bank, and row). Figure 2 illustrates the variability in the access latency by dictating example access scenarios that exhibit the effects from one or more of these factors. Figure 2 considers a sequence of two consecutive requests accessing DDRx. Using a sequence of two requests instead of just one request is necessary to show the case for delays due to the change of the DRAM state caused by commands of a previous request. The second request with subscript 1 in the figure is the request under consideration, while the first request (with subscript 0) is the previous request. Although we study all possible scenarios for both a read and a write request, due to space limitations, Figure 2 delineates command interactions for the considered request being a read. Scenarios for a write access are similar to the dictated ones. The chart in the upper right of Figure 2 delineates the access latency of the considered request for each scenario, while the table briefly explains the scenarios. As per Definition 1, access latency in Figure 2 is measured from the arrival of the considered request at the head of the queue at cycle 0, until the start of its data transfer ($\text{DATA}_1$). Complying with the majority of the commodity as well as predictable DRAM controllers, we assume that once the DRAM started executing one command of a request, it cannot be preempted, and commands from different requests can be pipelined to increase performance. The specific clock cycle values in Figure 2 reflects the timing constraints of DDR3-1600 as tabulated in Table I; however, the general scenarios apply for all DDRx devices since the basics of the access protocol remain the same.

*1) Targeting an Open Row:* Figures 2a–2e represent scenarios in which the considered request targets a row that is already open in the row buffer. Accordingly, the request does not need the precharge nor the activate stages, and only issues a R command. Figure 2a depicts the best case scenario: in addition to targeting an open row, $R_1$ satisfies all timing constraints upon arrival; hence, the controller issues $R_1$ immediately. DRAM takes $tRL$ cycles to place the data into the data bus. In Figure 2b, $R_1$ arrives directly after the memory has serviced another read, $R_0$, to a different rank; accordingly, $R_1$ has to be delayed by $BL/2 + tRTRS$ cycles. This is because data transfers from different ranks according to the standard has to be separated by the rank switching latency of $tRTRS$ cycles. Similarly, in Figure 2c, arrives directly after the memory has serviced another write, $W_0$. Therefore, $R_1$ has to be delayed by $tWL + BL/2 + tRTRS - tRL$ cycles.

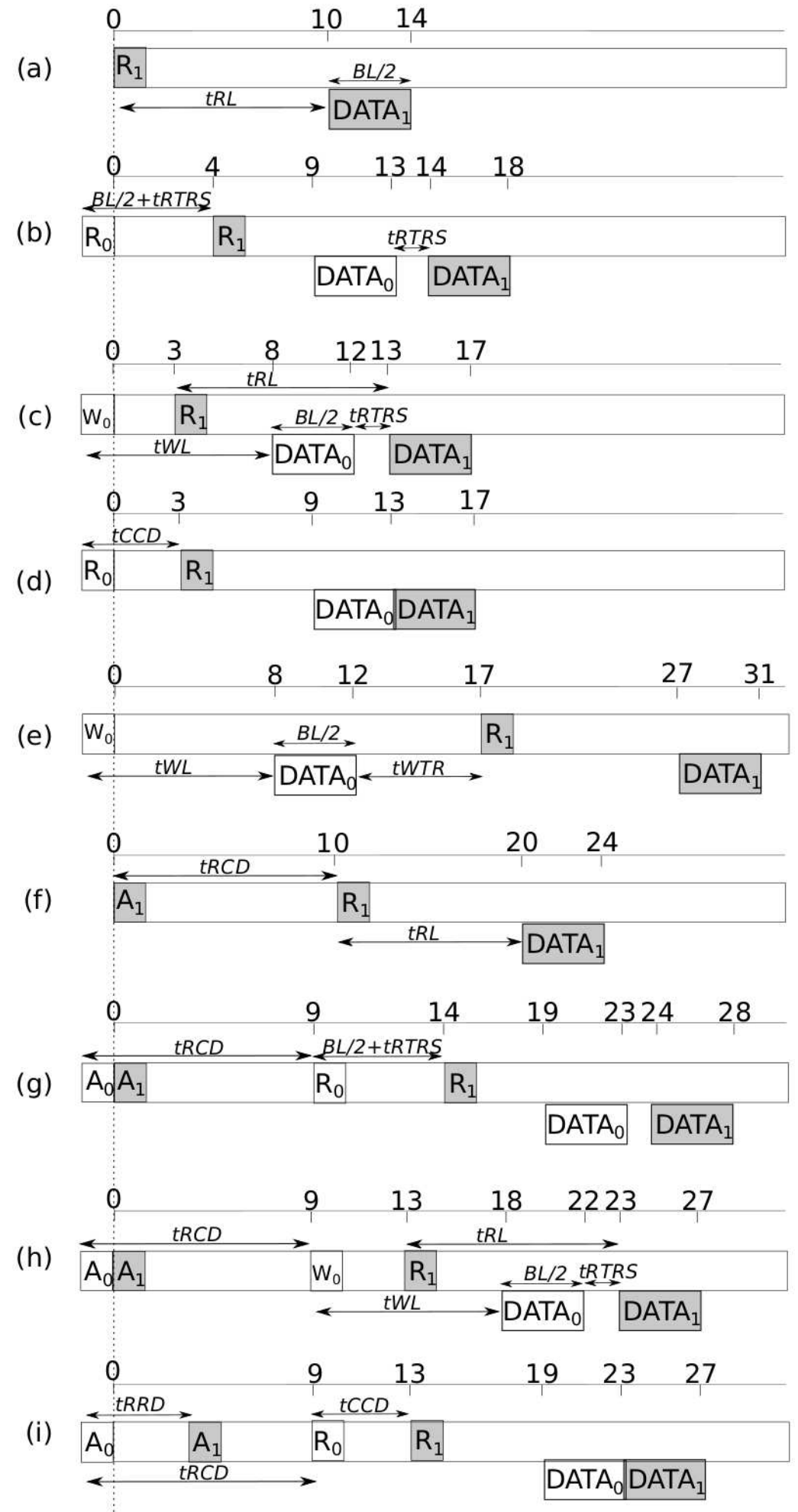

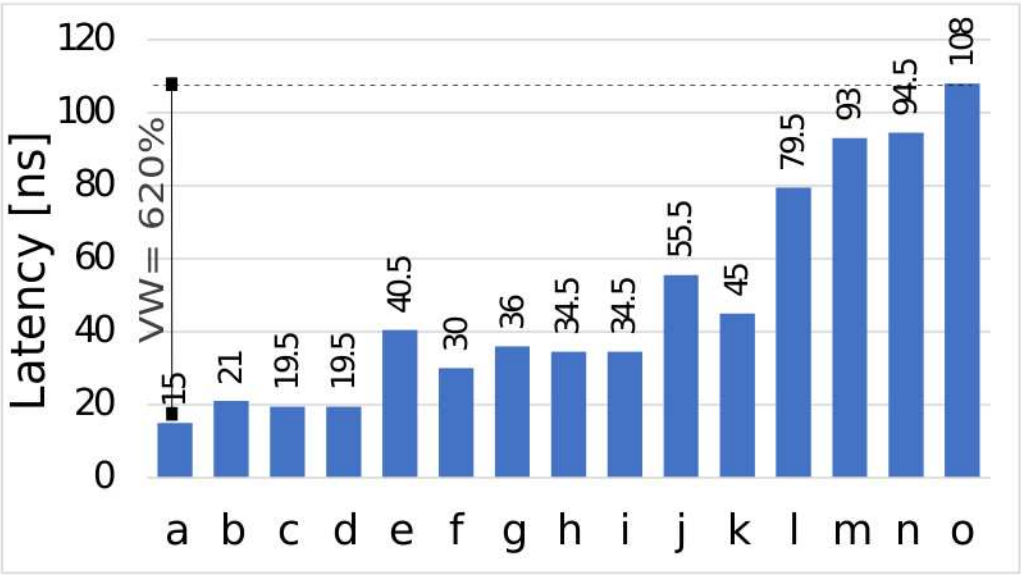

| | |
|---|---|
| (a)–(e) | Target an open row |
| (a) | $R_1$ satisfies all constraints |
| (b) | Arrives after a read access to a different rank |
| (c) | Arrives after a write access to a diffent rank |
| (d) | Arrives after a read access to same rank |
| (e) | Arrives after a write access to same rank |
| (f)–(j) | Target a closed row |
| (f) | $A_1$ satisfies all constraints |
| (g) | Arrives after a read access to a closed row in a different rank |
| (h) | Arrives after a write access to a closed row in a different rank |
| (i) | Arrives after a read access to a closed row in a different bank and same rank |
| (j) | Arrives after a write access to a closed row in a different bank and same rank |
| (k)–(o) | Target a conflict row |
| (k) | $P_1$ satisfies all constraints |
| (l) | Arrives after a read access to a closed row in same bank |
| (m) | Arrives after a write access to a closed row in same bank |
| (n) | Arrives after a read access to a conflict row in same bank |
| (o) | Arrives after a write access to a conflict row in same bank |

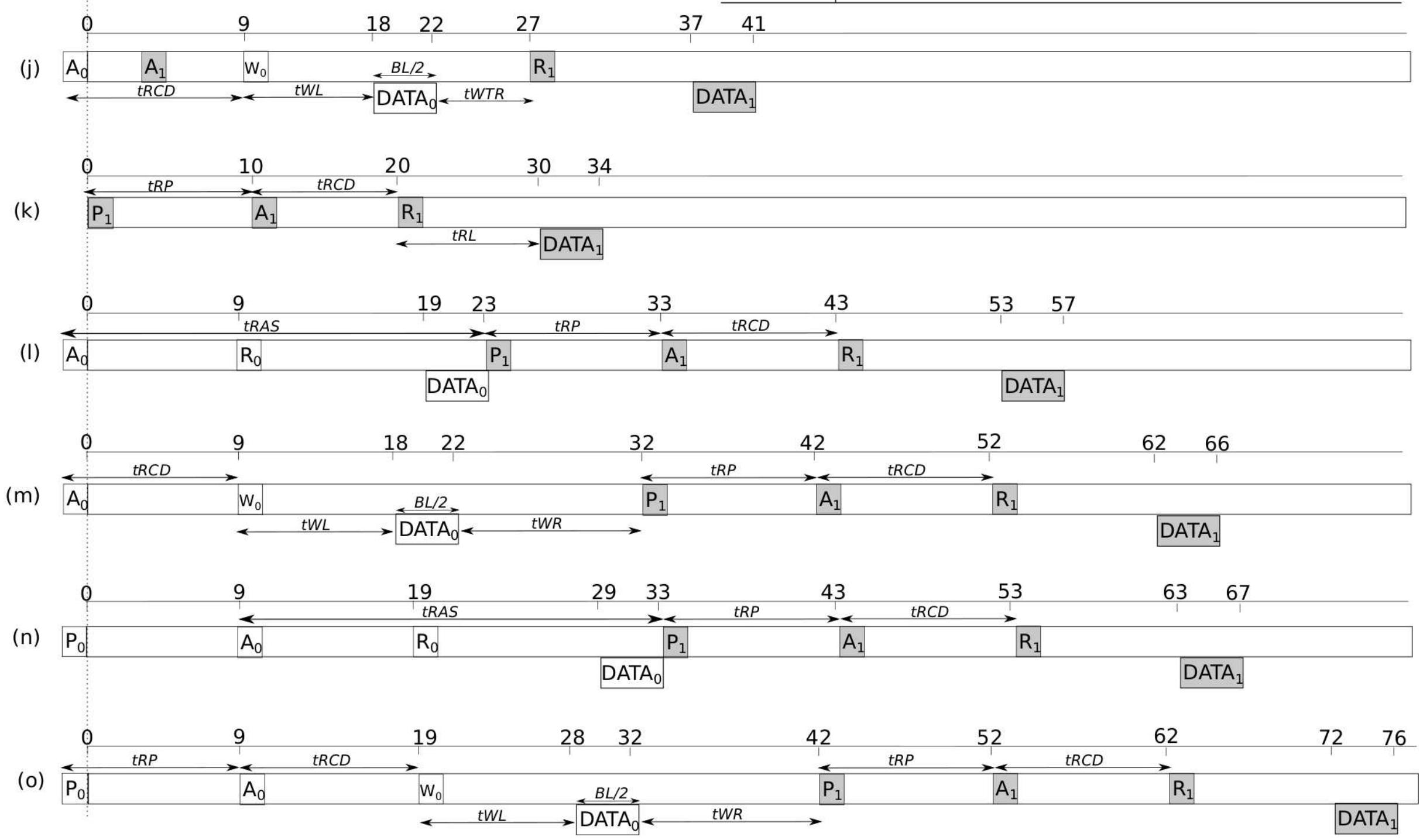

Fig. 2: Different DDRx access scenarios based on the arrival time of the request and the state of the memory from the directly previous request. The considered request is a read and arrives at the head of the queue at time 0. Timing constraints are relatively scaled based on JEDEC standard for DDR3-1600 [20] (Table I), with a clock of 1.5ns. Subscripts are for request numbers. Latency is measured from arrival to the start of data transfer (latency for (a) is 10 cycles or $10 \times 1.5 = 15ns$).

In Figure 2d, $R_1$ arrives directly after the memory has serviced $R_0$ to the same rank (either same bank or not). According to the JEDEC standard, it has to be delayed by $tCCD$ cycles before conducting the access. $R_1$ in Figure 2e is further delayed by the data bus turnaround time. Since the DRAM bus is bidirectional, certain delay has to elapse between every two successive requests of different types to same rank. $R_1$ arrives after a write request, $W_0$, to same rank; thus, it has to be delayed by a W-to-R delay of $tWL + BL/2 + tWTR$ cycles.

*2) Targeting a Closed Row:* In Figures 2f–2j, the considered request targets a closed row. Accordingly, it consists of $A_1$ and $R_1$ commands. In Figure 2f, the request arrives such that all timing constraints invoked due to previous requests are already satisfied. Therefore, the controller immediately issues $A_1$ at cycle 0, waits for the row to be activated ($tRCD$ cycles), and then issue the $R_1$ command at cycle 10. In Figures 2g and 2h, the request arrives while the memory is servicing a request to a different rank. $A_1$ is still issued immediately since the standard does not impose constraints among A commands to different ranks. However, $R_1$ has to wait for additional delays due to this previous request. In Figure 2g, it has to wait for $BL/2+tRTRS$ cycles after $R_0$ similar to Figure 2d, while in Figure 2h, it has to wait for $tWL+BL/2+tRTRS-tRL$ cycles after $W_0$ similar to Figure 2e. In Figures 2i and 2j, the request arrives while the memory is servicing a previous request to a different bank in same rank. In this case, $A_1$ and $A_0$ has to be separated by $tRRD$ cycles in addition to the aforementioned timing constraints related to $R_1$.

*3) Targeting a conflict Row:* Figures 2k–2o are for a request that targets a conflict row, and thus, consists of $P_1$, $A_1$, and $R_1$. In Figure 2k, the request satisfies all timing constraints upon arrival. The controller immediately issues $P_1$ at cycle 0 to precharge the existing row in the row buffer, an operation that consumes $tRP$ cycles. Afterwards, it issues the $A_1$ command to activate the row, and then issues $R_1$ after additional $tRCD$ cycles. In Figures 2l and 2m, the request arrives one cycle after the memory controller issued $A_0$ command of a previous request to a different row in the same bank. This previous request is a read in Figure 2l. Consequently, $P_1$ is delayed by $tRAS - 1$ cycles. On the other hand, in Figure 2m, the previous request is a write. Hence, $P_1$ cannot be issued before $tWR$ cycles after writing the data of that previous request. Figures 2n and 2o are similar to Figures 2l and 2m, respectively. However, the considered request arrives directly after a $P_0$ command instead of a $A_0$. Therefore, $P_1$ is further delayed by extra $tRP$ cycles.

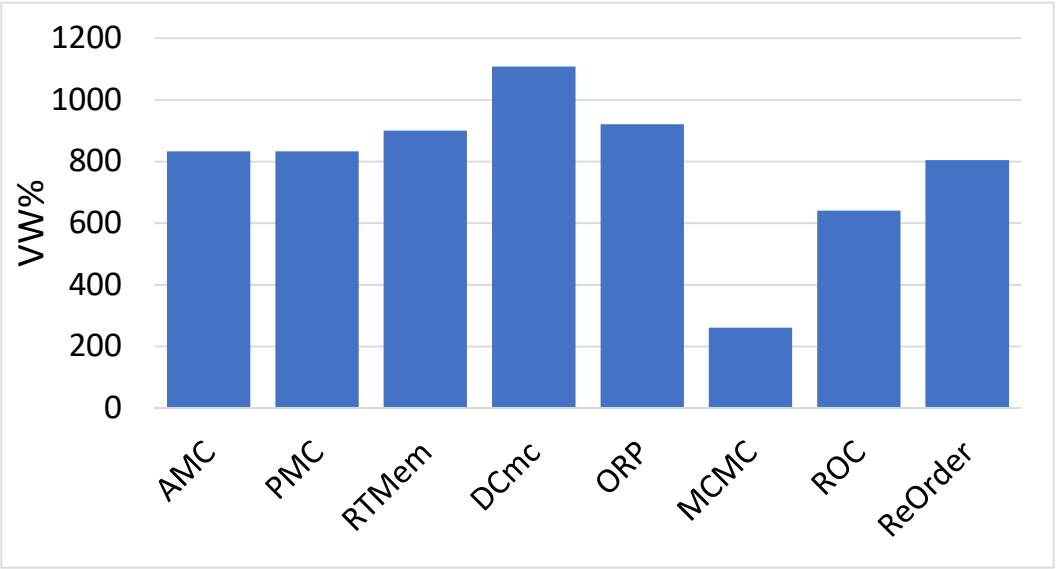

Fig. 3: Analytical variability window for different predictable memory controllers.

*4) Variability Window:* While Figure 2a represents the best-case scenario for the two-request sequence with the access latency of the considered request being 10 cycles, Figure 2o represents the worst-case scenario with access latency of 72 cycles. As the chart in Figure 2 highlights, the resulting variability window is 620%. Obviously, this is significantly high, which makes achieving predictability in DRAM a challenging task. Moreover, this value does not include the variability of the memory controller behavior. When considered, the variability window is expected to further increase since the memory controller scheduler accounts for large number of arbitration decisions. Examples include prioritizing requests from certain processing elements over the others, arbitrating based on transaction type (read vs write), and scheduling based on request addresses such as per-rank and per-bank arbitration. As we show in the next subsection, even state-of-the-art DDRx controllers aimed at real-time systems suffer significant variability in memory latency.

### C. Variability Window in Predictable Memory Controllers

We study both analytically and empirically the memory behavior of the state-of-the-art predictable memory controllers: AMC [15], PMC [12], RTMem [16], DCmc [14], ORP [17], MCMC [9], ROC [18], and ReOrder [7], [8]. In this section, we discuss the analytical results, while we discuss the empirical results in the evaluation section (Section VI).

Figure 3 delineates the variability window of these eight controllers. The variability window is calculated by Equation 1 using the analytical WCL and BCL of each controller. The WCL is based on the analysis provided in [26]. We assume a system with four PEs. For controllers aimed at multi-rank DRAMs (MCMC, ROC, and ReOrder), we calculate the VW for a 4-rank memory system. For controllers that require knowledge about the hit ratio (DCmc, ORP, ROC, and ReOrder), we assume a hit ratio of 35%. Hit ratio is the percentage of requests accessing a row that is already existing in the row buffer. Computing the WCL for different number of ranks or different hit ratios is not the focus of this paper and is already studied in the the corresponding papers of these controllers as well as in the comparative study in [26].

TABLE II: Timing constraints for RLDRAM3 [21].

| Parameter | Delay Description | Cycles |
|---|---|---|
| $tRC$ | Minimum time between two commands to same bank | 6 |
| $tWL$ | Minimum time between W to start of data transfer | 14 |
| $tRL$ | Minimum time between R to start of data transfer | 13 |
| $BL/2$ | Minimum time between two commands of same type | 4 (BL8) |
| $tRL - tWL + BL/2$ | Minimum time between R to W command | 3 (BL8) |
| $tWL - tRL + BL/2$ | Minimum time between W to R command | 5 (BL8) |

We compute the BCL assuming the considered request does not suffer from any additional delays due to other requests. In other words, it encounters only memory access latency. For detailed analysis of this section, we refer the interested reader to the technical report [32]. As Figure 3 illustrates, the variability window of these predicable controllers is huge. It exceeds $800\%$ in 6 out of the 8 studied controllers. We observe that controllers with multi-rank support (MCMC, ROC, and ReOrder) provide less variability window. For instance, MCMC has the least variability window of $261\%$. This is because these controllers mitigate the interference among different PEs by partitioning banks among PEs as well as mitigates the bus switching delays by alternating between different ranks. However, the is still large and can be ill-suited for real-time systems with tight safety-critical timing requirements.

As aforementioned, this high variability is due to the physically inherent limitations of the DDRx memories that induce large timing constraints, which all controllers have to satisfy. As a result, we believe that exploring other types of off-chip memories that address these limitations is unavoidable towards providing more predictable memory performance with less variability and tighter bounds.

## IV. RLDRAM for Real-Time Systems

RLDRAM is an emerging DRAM currently led by Micron [21] and provides a remarkable lower access latency compared to DDRx protocols. RLDRAM has a similar structure to DDRx as depicted in Figure 1. Nonetheless, RLDRAM achieves lower access latency by adopting unique architecture features that do not exist in commodity DDRx DRAMs. Two major features are of particular interest. First, RLDRAM uses an SRAM-like non-multiplexed address mode. All address bits are provided to the memory in one step as opposed to the two-step multiplexed mode in DDRx. Second, the row management through activation and precharging is handled internally by the RLDRAM device instead of the memory controller in case of DDRx. These two features together lead to multiple advantages of RLDRAM: 1) simplifying the access protocol, as accesses to RLDRAM consist of only R or W commands; 2) achieving low random access delay ($tRC$), which has a direct effect on worst-case access latency; and 3) decreasing bus turnaround (W-to-R and R-to-W) delays. Table II lists the most relevant timing constraints of RLDRAM3 running at 1600MHz. Overall, these advantages enable RLDRAM to provide a significant reduction in access latency, while incurring less variability as compared to DDRx memories as we illustrate in next subsection.

### A. RLDRAM Variability Window

Figure 4 delineates possible access scenarios of a request to RLDRAM similar to Figure 2 for DDRx. The chart in the bottom right shows the latency of the considered request for each scenario, while the table in the bottom left briefly explains each scenario. Unlike DDRx, RLDRAM is more deterministic, which explains the small number of access scenarios. Figure 4a (4b) depicts the best case scenario for a read (write) request that satisfies all timing constraints upon arrival. In Figures 4c and 4f, the considered request arrives at the head of the queue directly after the memory started servicing a previous request of the same type to a different bank. Thus, it is delayed by $BL/2 - 1$ cycles. In Figures 4d and 4e, the request arrives at the head of the queue directly after the memory started servicing a previous request of a different type to a different bank. Therefore, $W_1$ in Figure 4d has to be separated from the previous $R_0$ by a R-to-W delay of $tRL + BL/2 - tWL$. Similarly, $R_1$ in Figure 4e has to be separated from the previous $W_0$ by $tWL + BL/2 - tRL$, which is the W-to-R bus turnaround time. Finally, Figures 4g and 4h show the worst-case scenario, which is a bank conflict. The considered request arrives after the memory started servicing a request to the same bank; thus, it has to be delayed by $tRC - 1$ cycles. Since the type of the previous request is irrelevant in Figures 4g and 4h, $C_0$ indicates either a $R_0$ or $W_0$.

*1) Variability Window:* The scenario in Figure 4a incurs the $BCL$, which equals to 13 cycles. Contrarily, the scenario in Figure 4f encounters the $WCL$, which equals to 19 cycles. As a result, the variability window for the given sequence to the RLDRAM3-1600 is $46.2\%$ as the chart in Figure 4 delineates.

Compared to DDRx, RLDRAM provides $13.4\times$ reduction in the latency variability and $3.79\times$ reduction in the worst-case access latency. This identifies RLDRAM as promising solution towards providing a main memory with better predictable behavior and tighter WCL for real-time systems.

## V. Predictable RLDRAM Controller

To enable the usage of RLDRAM in real-time systems, we propose RLDC as a predictable RLDRAM memory controller that manages accesses to the RLDRAM. Figure 5 depicts the high-level architecture of RLDC. RLDC translates the memory requests into the corresponding RLDRAM commands and ensures the satisfaction of timing constraints amongst commands. It also predictably arbitrates amongst requests from PEs in a multi-processor system.

### A. Bank Partitioning vs Bank Sharing

Once a request is received by the memory controller, the Processor Decoder decodes the request's PE identification (Id) by using the PE bits encoded with the request. Then the Command Generation block generates the corresponding command by using the operation type of the request (read or write). Simultaneously, address translation is conducted by the Address Mapping block in Figure 5 to determine which bank, row, and column to access. The proposed controller allows for two different memory layouts: bank partitioning and bank sharing. Bank partitioning partitions RLDRAM banks across PEs, where each PE obtains an exclusive access to specific bank(s). On the other hand, bank sharing allows each PE to access any bank. Bank partitioning reduces the interference among PEs, while bank sharing provides more flexibility. The user selects the layout through one bit in a memory configuration register inside the controller (not shown in Figure 5). We conduct latency analysis for both mechanisms and experimentally compare their behaviors. For bank partitioning, the Address Mapping conducts the partitioning based on the Id provided by the Processor Decoder. Once a request's command is generated and its bank is calculated, this information is buffered in the corresponding PE queue to be scheduled by the predictable arbiter.

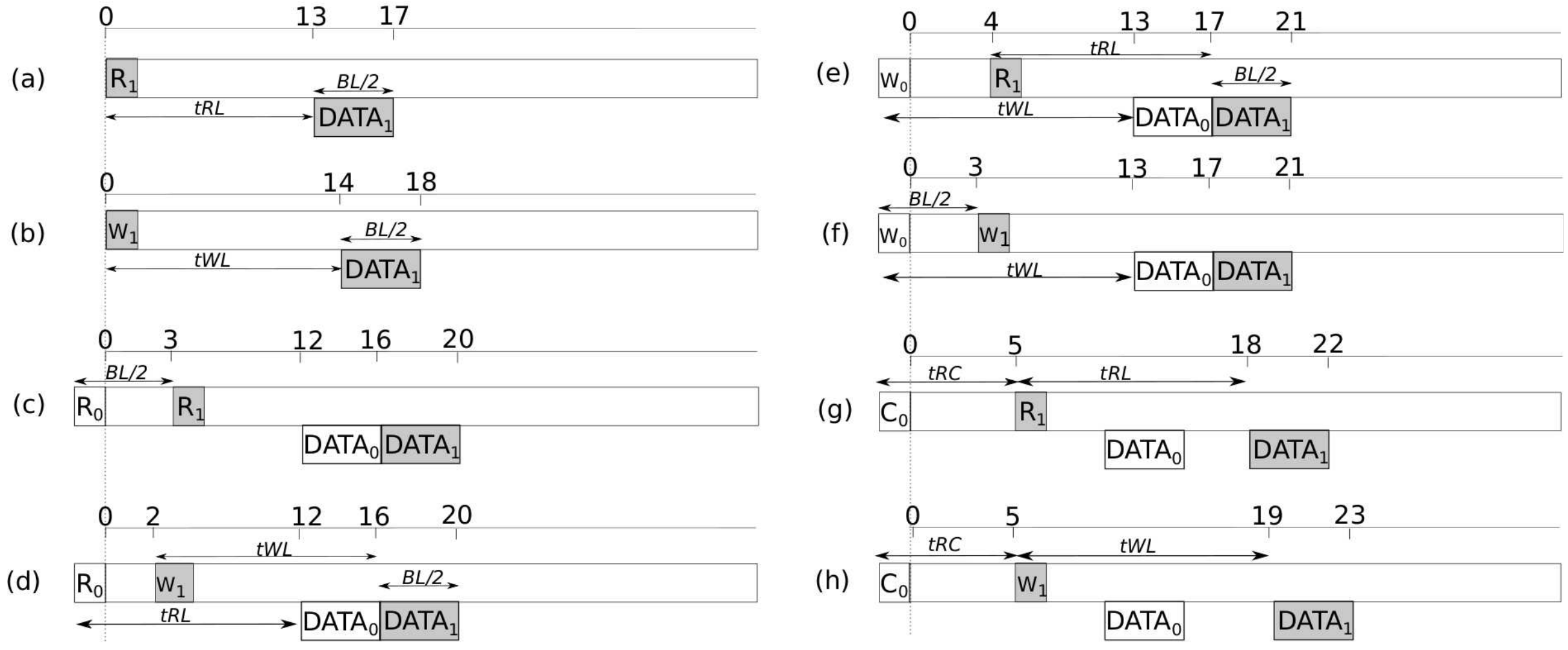

| | |
|---|---|
| (a) | $R_1$ satisfies all constraints |
| (b) | $W_1$ satisfies all constraints |
| (c) | $R_1$ Arrives after a read access to a different bank |
| (d) | $W_1$ Arrives after a read access to a different bank |
| (e) | $R_1$ Arrives after a write access to a different bank |
| (f) | $W_1$ Arrives after a write access to a different bank |
| (g) | $R_1$ Arrives after a read or write access to same bank |
| (h) | $W_1$ Arrives after a read or write access to same bank |

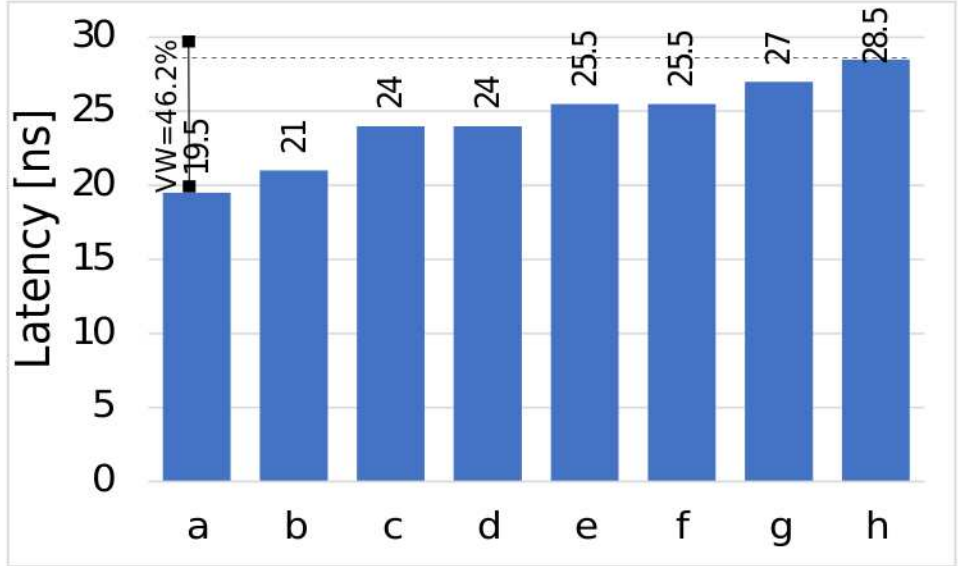

Fig. 4: Different RLDRAM access scenarios. Timing constraints are relatively scaled based on the timing constraints of RLDRAM3-1600 with a clock of 1.5ns (Table II). Subscripts are for request numbers.

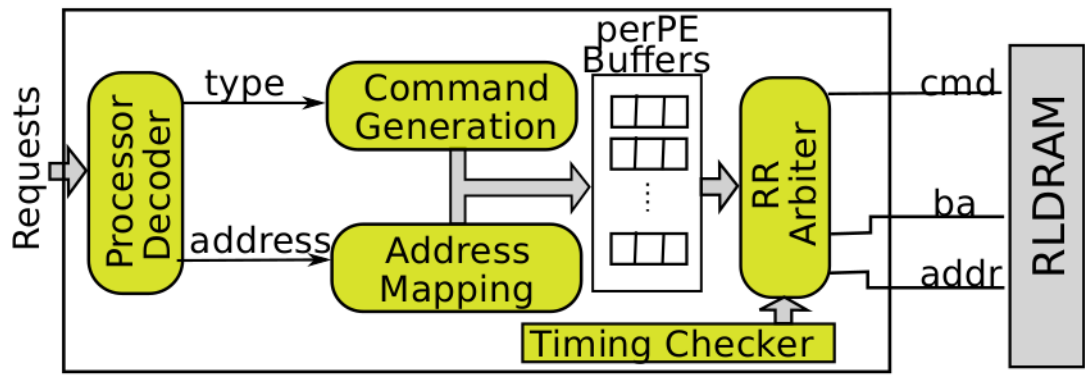

Fig. 5: High level architecture of RLDC.

### B. Predictable Arbitration

The proposed controller deploys Round Robin (RR) arbitration (RR Arbiter in Figure 5) amongst requests at the head of each processor buffer (perPE Buffers). RR is a dynamic predictable arbitration mechanism that facilitates the latency analysis without sacrificing average-case performance. At the beginning of each cycle, the arbiter checks if the PE with the current slot in the RR schedule has a ready request to be sent to the RLDRAM. Timing Checker block decides if the request at the head of the queue of this processor satisfies the timing constraints of the RLDRAM and can be immediately serviced. This is conducted by maintaining a counter for each timing constraint. To exemplify, if RLDC issued a R to a bank, the $tRC$ counter of that bank is initialized by the $tRC$ constraint value. Hence, the Timing Checker ensures that no other command is issued to that bank before the $tRC$ counter reaches zero. If the request is ready, the arbiter issues it to the RLDRAM in the form of a command (either R or W), a bank address, and request address. These are the cmd, ba, and addr signals in Figure 5 at the interface between the controller and the RLDRAM. If the request is not ready due to timing violations, the arbiter checks the next PE in the schedule.

### C. Latency Analysis

We derive both worst- and best-case values for the total memory latency (Definition 2) incurred by any request to the proposed RLDRAM solution. The analysis is conducted for a multi-processor system. Although the proposed solution works for any pipeline architecture, we conduct the analysis assuming in-order PEs as they better represent PEs used for real-time systems. In addition, it is the commonly assumed pipeline type in predictable DDRx solutions including the ones we compare against (e.g. [12], [17]). For sake of generality, we derive the worst-case total latency for the two supported memory layouts: 1) bank sharing, where any PE has access to all banks (Lemma 1), and 2) bank partitioning, where banks are privately assigned to PEs (Lemma 2). Moreover, to enable the analytical calculation of the variability window, Lemma 3 provides the best-case total latency, which is the same for both bank partitioning and bank sharing mechanisms.

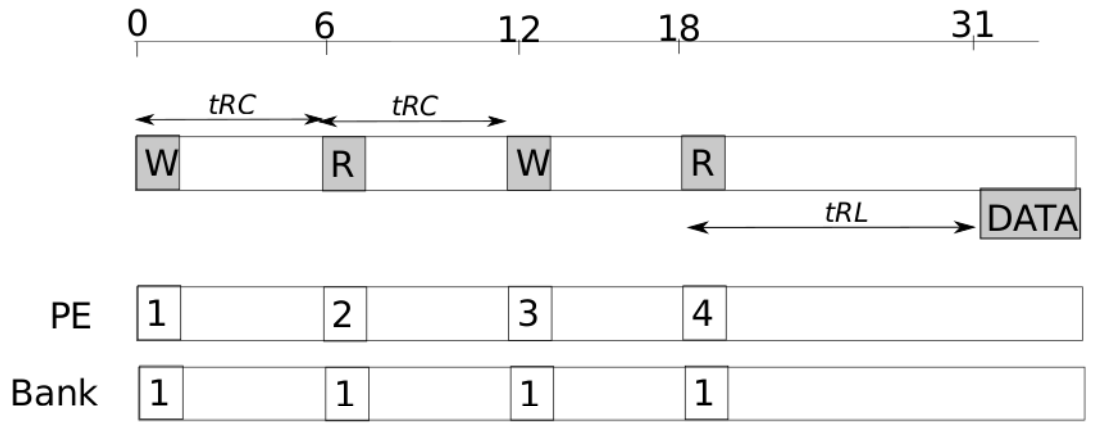

Fig. 6: WCL in a four-PE sysem with bank-sharing RLDC.

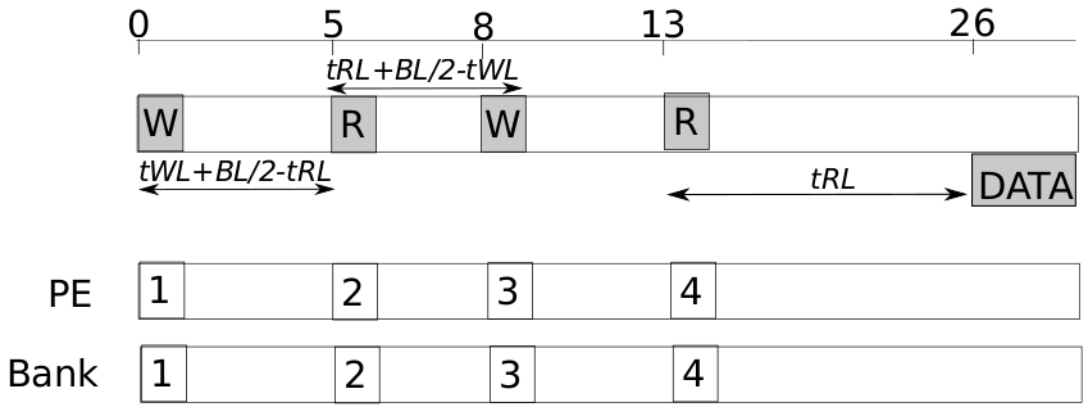

Fig. 7: WCL in a four-PE sysem with bank-partitioning RLDC.

*Lemma 1:* The worst-case total latency of a request $Req_i$ from $PE_i$ in a system with $N$ PEs and a bank sharing layout can be calculated as follows (where $tCL$ is $tRL$ if $Req_i$ is a read and $tWL$ if $Req_i$ is a write):

$$WCL_i^{share} = (N-1) \times tRC + tCL.$$

*Proof:* Recall that RLDC implements RR arbitration among PEs. As a consequence, $PE_i$ in the worst case waits for all other $N-1$ PEs before it is granted access. Additionally, since bank partitioning is not deployed, requests from different PEs can target any bank. In the worst case, requests from all PEs target the same bank such that only one command is serviced every $tRC$ cycles. Accordingly, $PE_i$ has to wait for $(N-1) \times tRC$ cycles before it gains access to the RLDRAM. Once the command of $Req_i$ is issued to the memory (R or W), the data transfer will start after $tRL$ or $tWL$ with respect to the type. Figure 6 delineates this scenario for $N = 4$, where the WCL is $3 \cdot tRC + tRL = 3 \cdot 6 + 13 = 31$ cycles. ■

*Lemma 2:* The worst-case total latency of a request $Req_i$ from $PE_i$ in a system with $N$ PEs and a bank partitioning layout can be calculated as:

$$WCL_i^{part} = \lceil \frac{N-1}{2} \rceil \times (tWL - tRL + BL/2) + \lfloor \frac{N-1}{2} \rfloor \times (tRL - tWL + BL/2) + tCL.$$

*Proof:* From proof of Lemma 1, $PE_i$ in the worst case waits for all other $N-1$ PEs before it is granted access. Since bank partitioning is deployed, requests from PEs are guaranteed to access different banks (assuming that $N$ is less than the number of banks). Accordingly, the only constraint is to separate every two successive commands in the RR schedule by the minimum delay required to avoid data bus collisions. Three cases are possible for any two successive commands to RLDRAM. 1) Both commands are of same type (either R or W). In this case, the minimum delay between these two commands is $BL/2$. Figures 4c and 4f represent this case. 2) The two commands are a write followed by a read. In this case, the minimum W-to-R delay is $tWL - tRL + BL/2$. Figure 4e represent this case. 3) The two commands are a read followed by a write. From Figure 4d, these two commands have to be separated by a R-to-W delay of $tRL - tWL + tBUS$ cycles.

In the worst case, a data bus switching occurs between every two successive requests. Furthermore, since $tWL$ is larger than $tRL$ (Table II), the W-to-R delay is larger than the R-to-W delay. Thus, the worst-case number of W-to-R switches is equal to or larger than R-to-W switches, which justifies the ceiling and flooring operations in Lemma 2. Figure 7 delineates this worst case scenario for $N = 4$, which equals $2 \cdot (tWL - tRL + BL/2) + tRL - tWL + BL/2 + tRL = 2 \cdot 5 + 3 + 13 = 26$ cycles. ■

*Lemma 3:* The best-case total latency of a request $Req_i$ from $PE_i$, $BCL_i$, in a system with $N$ PEs is calculated as:

$$BCL_i = tCL.$$

*Proof:* In best case, $Req_i$ does not suffer any interference latency from other requests. Accordingly, its command is ready to execute upon arrival. Since there is a minimum of $tRL$ ($tWL$) cycles between the R (W) command and the start of its data transfer, $BCL$ is as calculated in Lemma 3. ■

From Lemmas 1-3, the variability windows for bank sharing and bank partitioning RLDC are calculated in Equations 2 and 3, respectively.

$$VW^{share} = \frac{(N-1) \times tRC}{tCL} \times 100 \quad (2)$$

$$VW^{part} = \big(\lceil \frac{N-1}{2} \rceil \times (tWL - tRL + BL/2) + \lfloor \frac{N-1}{2} \rfloor \times (tRL - tWL + BL/2)\big) \times \frac{100}{tCL} \quad (3)$$

## VI. Evaluation

To evaluate the effectiveness of the proposed predictable RLDRAM solution, we use MacSim [33], a multi-processor architectural simulator integrated with DRAMSim2 [34] as the main memory system. We extend DRAMSim2 to faithfully model the RLDRAM operation and implement the proposed RLDC to manage accesses to the RLDRAM. We compare the proposed solution with eight of the state-of-the-art predictable DDRx controllers: AMC [15], PMC [12], [13], RTMem [16], DCmc [14], ORP [17], MCMC [9], ROC [18], and ReOrder [7], [8]. On the integration of these controllers, we reuse the open-source implementation provided by [26], [35]. Table III tabulates the important system information. We use Benchmarks from the EEMBC-auto suite [36], which includes representative applications from the embedded automotive domain. We use *a2time* benchmark as the application under analysis running on one PE, while the three other PEs are executing interfering applications. For these interfering PEs, we pick the three most memory extensive benchmarks from the EEMBC-auto suite: *matrix, aifftr, and aiifft*.

TABLE III: Simulation environment configurations.

| | |
|---|---|
| PEs | 4 PEs, in-order pipeline, a private 16KB L1 and a shared 1MB L2 cache |
| Main Memory | Either RLDRAM or DDR |
| RLDRAM | RLDRAM3-1600 [21] with timing constraints in Table II, while the proposed RLDC manages accesses to RLDRAM |
| DDRx | DDR3-1600 with timing constraints with timing constraints in Table I, while AMC, PMC, RT-Mem, DCmc, ORP, MCMC, ROC, or ReOrder manages access to DDR3 |
| Bank Management | We experiment with both bank partitioning and bank sharing among PEs for RLDC |

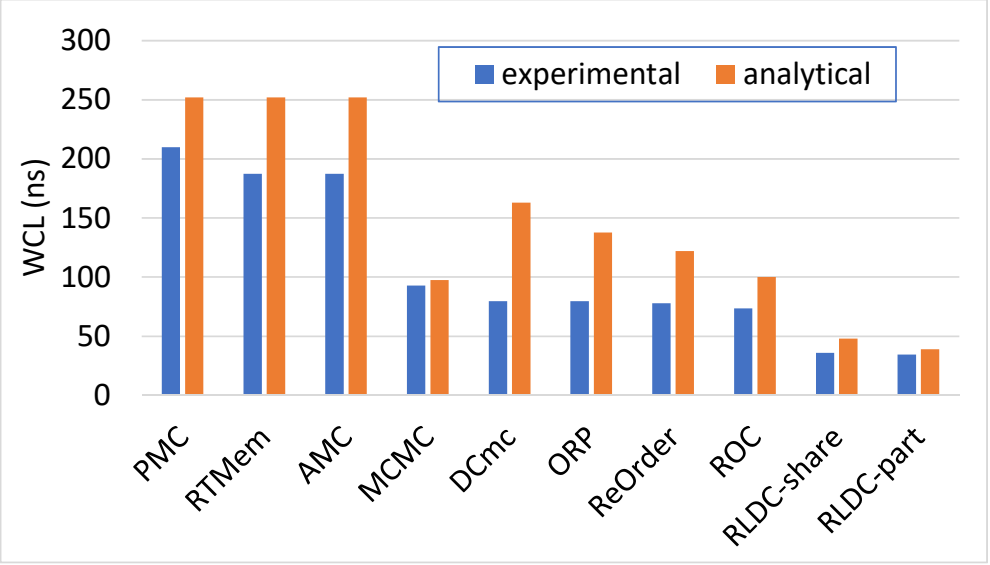

Fig. 8: Worst-case experimental and analytical latencies (ordered ascendingly by experimental WCL).

### A. *Worst-Case Latency*

Figure 8 delineates both the experimental and the analytical WCLs for the DDRx and RLDRAM systems used in the experiments. Experimental WCL is the maximum total latency suffered by a request from the PE under analysis to the main memory. It is measured from the arrival time instance of the request into the controller to the time instance when the corresponding data of this request start transferring on the data bus. Analytical WCL of RLDRAM is the latency bound derived by the timing analysis conducted in Section V-C. For DDRx, we use the latency bounds derived in Section III-C. RLDC-part in Figure 8 indicates that RLDC is configured to use the bank partitioning mechanism, while RLDC-share indicates RLDC with a bank sharing mechanism.

***Observations.*** 1) All experimental WCLs are less than their corresponding bounds, which confirms that the derived bounds are safe. 2) Clearly, the proposed RLDRAM solution provides a considerable less WCL compared to DDRx both experimentally and analytically. For instance, under bank partitioning mechanism, RLDC provides a WCL bound of $39ns$. On the other hand, WCL of DDRx varies from $97.5ns$ for MCMC to $252ns$ for PMC, RTMem, and AMC controllers. This is $2.5\times$ and $6.46\times$ higher than RLDC's WCL, respectively. Similar results are observed experimentally, WCL of RLDC with bank partitioning is $34.5ns$. Minimum DDRx WCL of $73.5ns$ is observed for ROC ($2.13\times$ RLDC's WCL), while the maximum WCL is observed for PMC and equals to $210ns$ ($6.09\times$ RLDC's WCL). It is worth noting that this relatively low WCL of MCMC, ROC, and ReOrder as compared to other DDRx controllers relies on the existence of four DDRx ranks in the system. For a single-rank DDRx, those multi-rank controllers lose this advantage. 3) RLDC with bank partitioning provides tighter WCLs compared to bank sharing. This is because bank partitioning allows each PE to obtain an exclusive access to specific bank(s), which reduces the interference. This comes at the expense of lack of flexibility. For instance, unlike bank sharing, partitioning does not allow data sharing between PEs. 4) The gap between the analytical and experimental WCL is excessively higher for most of the DDRx controllers. This is because DDRx has larger number of commands and timing constraints between them. This complexity of the DDRx leads to nondeterministic behavior with wide variability window, which we study in the next experiment in more details.

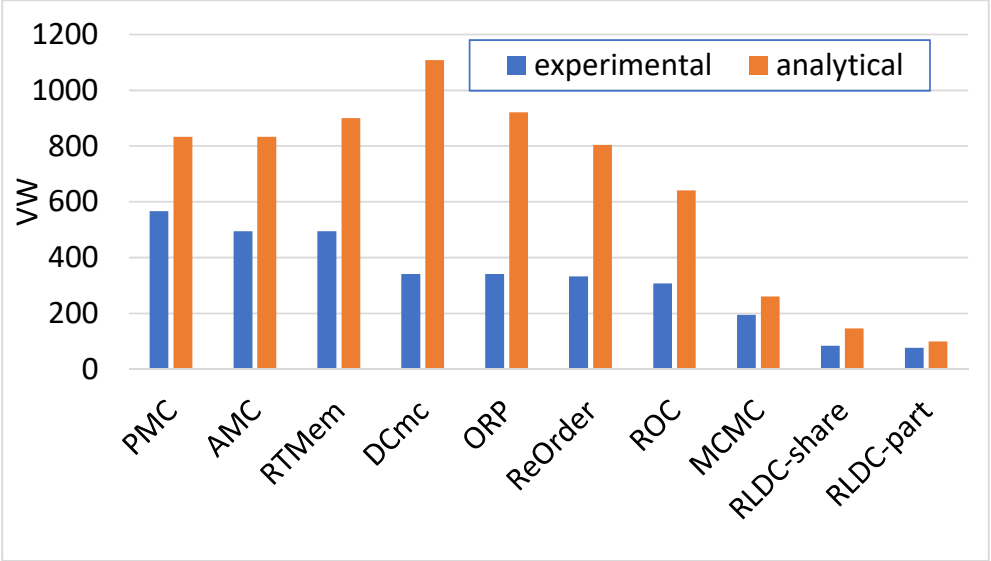

Fig. 9: Variability window (ordered ascendingly by experiemntal VW).

### B. *Variability in Total Request Latency*

Figure 9 plots the experimentally observed variability window in the total memory latency for RLDRAM and DDRx using the same setup as in Section VI-A. The experimental variability window for each controller is calculated based on the observed best- and worst-case total latencies of this controller. For sake of comparison, we also plot the analytical variability window from Section III-C.

***Observations.*** 1) Results show that a request to DDRx suffers from a significant variability in its latency and the variability window (Definition 3) is above $300\%$ for seven of the eight considered DDRx controllers. The eighth controller (MCMC) has a variability window of $195.2\%$.

2) Contrarily, RLDC provides a considerable reduction in the variability window: $84.6\%$ for bank sharing and $76.9\%$ for bank partitioning. The explanation for this is that variability window is the relative difference between BCL and WCL. BCL occurs when the request suffers no interference at all form other requests. So its command execute immediately upon arrival. For both RLDRAM and DDRx, a request in best case consists of a single command (R or W). Accordingly, the BCL is either $tRL$ or $tWL$ for a read or write request, respectively. Since these two constraints are less in DDR3 than in RLDRAM3 (From Tables I and II), DDR3 in fact has less BCL. On the other hand, because of the complexity of the DDRx protocol, WCL for DDRx is larger than that of RLDRAM as observed in Section VI-A. Accordingly, the variability window of DDRx is expected to be larger than RLDRAM3. This motivates the adoption of RLDRAM3 in real-time systems with its lower WCL and less variability.

### C. Scalability: Sensitivity to Number of Interfering PEs

In this experiment, we study the effect of varying the number of PEs in the system on the analytical memory latency bounds. Figure 10 depicts our findings.

***Observations.*** 1) Increasing the number of PEs, the gap in latency between RLDC and the majority of the considered DDRx controllers immensely increases. This is mainly because of the timing constraints that dominate the latency bounds, which also reflects the physical limitations of the DDRx memories. As explained in Section IV, RLDRAM does not suffer from the high latency of activation and precharging stages; thus, it has a lower $tRC$ delay. In addition, accesses to RLDRAM does not suffer from the high data bus switching penalties that exist in DDRx memories as the switching delay in RLDRAM is only one cycle. 2) Some DDRx controllers show better scalability (less increase in WCL with the increase in number of PEs) than others. DDRx controllers in Figure 10 can be classified into three categories. a) Controllers with bank sharing mechanisms: AMC, PMC, and RTMem, which suffer the maximum increase in WCL when increasing the number of PEs. b) Controllers with bank partitioning and multi-rank support: ReOrder, ROC, and MCMC, which incur the least WCL increase. This is because these controllers reduce interference among PEs by combining two techniques. First, they partition banks to eliminate the row conflict interference, which mitigates the long $tRC$ delay. Second, they use the multi-rank support to amortize the data bus switching delays. c) Controllers with only bank partitioning: ORP and DCmc, which exhibit intermediate increase in WCL. 3) Comparing all the aforementioned categories of DDRx controllers with RLDC (including the ones with bank partitioning and multi-rank requirement), highlights the advantages of RLDRAM for real-time systems. RLDC (whether with bank partitioning or sharing) encounters the least WCL across all PEs. This means that using RLDC, real-time systems enjoy tighter WCL with less sensitivity to the number of interfering PEs, while having the flexibility of sharing data among different PEs.

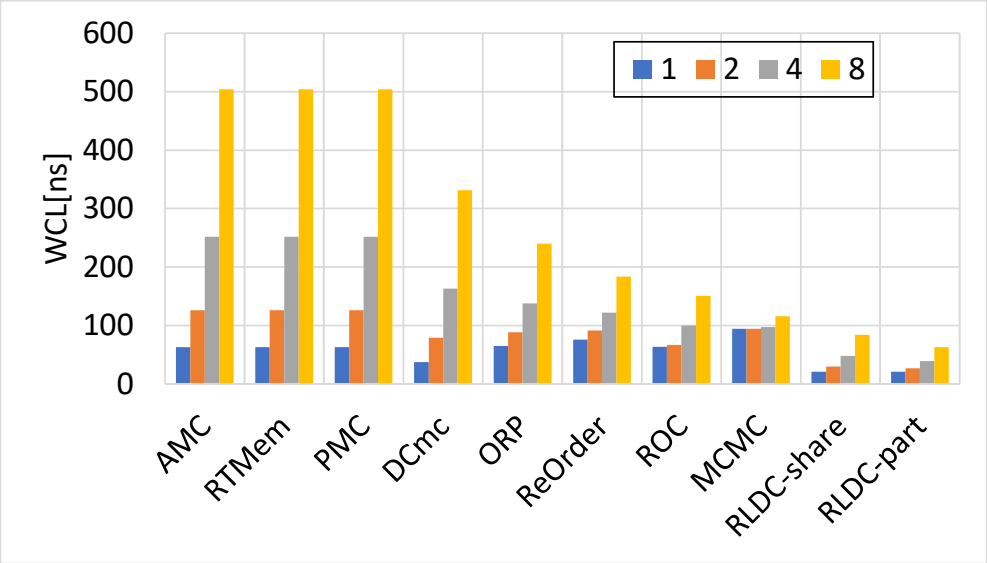

Fig. 10: Effect of varying number of PEs in the system.

## VII. Other Considerations: A discussion

We discuss some of the practical considerations towards adopting RLDRAM in real-time systems.

**Cost.** Compared to DDRx DRAMs, Static RAMs (SRAMs) provide a significantly lower latency at the expense of a significantly higher cost. This high cost prohibits the deployment of SRAMs in systems that require large capacity, which leaves the high-latency DRAMs as the only possible option. RLDRAM addresses this challenge by offering a balanced solution that provides a comparable latency to SRAMs, with a comparable cost to DDRx [27]. The lower latency is achieved by the means explained in Section IV. On the other hand, the lower cost than SRAMs is achieved by preserving the internal structure of DDRx, which consists of a single transistor as opposed to 6 in the case of SRAMs.

**Density.** Currently, the maximum density supported by RLDRAM3 is $2.25GB$ [27], while DDR3 offers up to $8GB$. Nonetheless, for real-time systems that require more than $2.25GB$ of data, multiple RLDRAM channels may be used.

**Adoption.** RLDRAM is manufactured by Micron [27], one of the biggest suppliers of memory devices. This ensures its stability and future adoption. It is also already adopted in high-speed networking solutions [27], [28]. Moreover, RLDRAM is supported by several industry players such as Intel [37], Xilinx [38], Lattice [39], and Northwest Logic [40]. As a result, state-of-the-art FPGA-based boards support RLDRAM interfacing (e.g. Intel Arria 10 GX FPGA [37] and Xilinx Virtex UltraScale VCU110 [38] development kits). We believe that this support is an appealing opportunity since it provides the necessary means to design, experiment and evaluate RLDRAM solutions for future real-time systems.

## VIII. Conclusions

The real-time community has been focusing on DDRx DRAMs as a sole solution for off-chip memories in real-time systems. We highlight the limitations of DDRx memories towards providing predictability in these systems. Then, we show that the emerging RLDRAM memory provides a promising solution that meets real-time requirements with tighter latency bounds and less variability. To enable this deployment, we provide a predictable RLDRAM memory controller supporting multi-processor systems and conduct timing analysis to bound the latency suffered by any memory request. We compare the proposed solution with competitive predictable DDRx controllers. Results show that the proposed RLDRAM solution provides up to $6.4\times$ reduction in the worst case memory latency and up to $11\times$ less latency variability.